\makeatletter
\def\ssize{\scriptstyle}
\newdimen\ex@
\def\vspace@{\def\vspace##1{\noalign{\vskip##1\relax}}}
\def\Let@{\relax\iffalse{\fi\let\\=\cr\iffalse}\fi}

\let\ampersand@\relax
\newdimen\bigaw@
\newdimen\minaw@
\newdimen\minCDaw@
\newif\ifCD@
\def\minCDarrowwidth#1{\relax\ifmmode\ifinner\onlydmatherr@\minCDarrowwidth
 \else\minCDaw@#1\relax\fi\else\onlydmatherr@\minCDarrowwidth\fi}
\def\CD{\bgroup\catcode`\@=\active 
 \vspace@\relax\let\ampersand@&\iffalse}\fi
 \CD@true\vcenter\bgroup\Let@\tabskip\z@skip\baselineskip20\ex@
 &\hfill$\m@th##$\hfill\cr}
\def\endCD{\cr\egroup\egroup\egroup}
\def\cd#1{\csname cd\string#1\endcsname}
\def\cddef#1{\expandafter\def\csname cd\string#1\endcsname}
\cddef.{\relax\ifmmode&&\else\leavevmode.\spacefactor3000 \fi}
\cddef\rightarrow#1#2{\harrow@{#1}{#2}\rightarrowfill} \cddef>{\cd\rightarrow}
\cddef\leftarrow#1#2{\harrow@{#1}{#2}\leftarrowfill} \cddef<{\cd\leftarrow}
\def\harrow@#1#2#3{\ampersand@\setbox\z@\hbox{$\ssize
 \;\;{#1}\;$}\setbox\@ne\hbox{$\ssize\;\;{#2}\;$}\setbox\tw@
 \hbox{$#2$}\ifCD@
 \global\bigaw@\minCDaw@\else\global\bigaw@\minaw@\fi
 \ifdim\wd\z@>\bigaw@\global\bigaw@\wd\z@\fi
 \ifdim\wd\@ne>\bigaw@\global\bigaw@\wd\@ne\fi
 \ifCD@\hskip.5em\fi
 \ifdim\wd\tw@>\z@
 \mathrel{\mathop{\hbox to\bigaw@{#3}}\limits^{#1}_{#2}}\else
 \mathrel{\mathop{\hbox to\bigaw@{#3}}\limits^{#1}}\fi
 \ifCD@\hskip.5em\fi\ampersand@}
\cddef\uparrow#1#2{\llap{$\vcenter{\hbox
 {$\ssize#1$}}$}\Big\uparrow\rlap{$\vcenter{\hbox{$\ssize#2$}}$}&&}
\cddef\downarrow#1#2{\llap{$\vcenter{\hbox
 {$\ssize#1$}}$}\Big\downarrow\rlap{$\vcenter{\hbox{$\ssize#2$}}$}&&}
\cddef=#1#2{\ampersand@\setbox\z@\hbox{$\ssize
 \;{#1}\;\;$}\setbox\@ne\hbox{$\ssize\;{#2}\;\;$}\setbox\tw@
 \hbox{$#2$}\ifCD@
 \global\bigaw@\minCDaw@\else\global\bigaw@\minaw@\fi
 \ifdim\wd\z@>\bigaw@\global\bigaw@\wd\z@\fi
 \ifdim\wd\@ne>\bigaw@\global\bigaw@\wd\@ne\fi
 \ifCD@\hskip.5em\fi
 \ifdim\wd\tw@>\z@
 \mathrel{\mathop{\vbox{\hrule width\bigaw@\vskip3\ex@\hrule width
 \bigaw@}}\limits^{#1}_{#2}}\else
 \mathrel{\mathop{\vbox{\hrule width\bigaw@\vskip3\ex@\hrule width
 \bigaw@}}\limits^{#1}}\fi
 \ifCD@\hskip.5em\fi\ampersand@}
\cddef|#1#2{\llap{$\vcenter{\hbox
 {$\ssize#1$}}$}\Big\vert\rlap{$\vcenter{\hbox{$\ssize#2$}}$}&&}
\cddef\|#1#2{\llap{$\vcenter{\hbox
 {$\ssize#1$}}$}\Big\vert\Big\vert\rlap{$\vcenter{\hbox{$\ssize#2$}}$}&&}
\def\pretend#1\haswidth#2{\setbox\z@\hbox{$\scriptstyle{#2}$}\hbox
 to\wd\z@{\hfill$\scriptstyle{#1}$\hfill}}
\cddef\updownarrows#1#2{\llap{$\vcenter{\hbox{$\ssize#1$}}$}
 \Big\uparrow\Big\downarrow\rlap{$\vcenter{\hbox{$\ssize#2$}}$}&&}
\cddef\downuparrows#1#2{\llap{$\vcenter{\hbox{$\ssize#1$}}$}
 \Big\downarrow\Big\uparrow\rlap{$\vcenter{\hbox{$\ssize#2$}}$}&&}
\cddef\rightleftarrows#1#2{\harrows@{#1}{#2}\rightarrowfill\leftarrowfill}
\cddef\leftrightarrows#1#2{\harrows@{#1}{#2}\leftarrowfill\rightarrowfill}
\def\harrows@#1#2#3#4{\ampersand@\setbox\z@\hbox{$\ssize
 \;{#1}\;\;$}\setbox\@ne\hbox{$\ssize\;{#2}\;\;$}\setbox\tw@
 \hbox{$#2$}\ifCD@
 \global\bigaw@\minCDaw@\else\global\bigaw@\minaw@\fi
 \ifdim\wd\z@>\bigaw@\global\bigaw@\wd\z@\fi
 \ifdim\wd\@ne>\bigaw@\global\bigaw@\wd\@ne\fi
 \ifCD@\hskip.5em\fi
 \ifdim\wd\tw@>\z@
 \mathrel{\mathop{\harrowsfill@#3#4}\limits^{#1}_{#2}}\else
 \mathrel{\mathop{\harrowsfill@#3#4}\limits^{#1}}\fi
 \ifCD@\hskip.5em\fi\ampersand@}
\def\harrowsfill@#1#2{\stackrel{\raisebox{0pt}[2\ex@][0pt]{\hbox
 to\bigaw@{#1}}}{\raisebox{-1\ex@}[0pt][0pt]{\hbox to \bigaw@{#2}}}}

\def\err@#1{\errmessage{AmS-TeX error: #1}}
\newhelp\athelp@
{Only certain combinations beginning with @ make sense to me.^^J
Perhaps you wanted \string\@\space for a printed @?^^J
I've ignored the character or group after @.}
\def\futureletnextat@{\futurelet\next\at@}
{\catcode`\@=\active
\lccode`\Z=`\@ \lccode`\I=`\I \lowercase
{\gdef@{\expandafter\csname futureletnextatZ\endcsname}\expandafter
 \gdef\csname atZ\endcsname
 {\ifcat\noexpand\next a\def\next{\csname atZZ\endcsname}\else
 \ifcat\noexpand\next0\def\next{\csname atZZ\endcsname}\else
 \ifcat\noexpand\next\relax\def\next{\csname atZZZ\endcsname}\else
 \def\next{\csname atZZZZ\endcsname}\fi\fi\fi\next}
\expandafter\gdef\csname atZZ\endcsname#1{\expandafter
 \ifx\csname #1Zat\endcsname\relax\def\next
 {\errhelp\expandafter=\csname athelpZ\endcsname
 \csname errZ\endcsname{Invalid use of \string@}}\else
 \def\next{\csname #1Zat\endcsname}\fi\next}
\expandafter\gdef\csname atZZZ\endcsname#1{\expandafter
 \ifx\csname \string#1ZZat\endcsname\relax\def\next
 {\errhelp\expandafter=\csname athelpZ\endcsname
 \csname errZ\endcsname{Invalid use of \string@}}\else
 \def\next{\csname \string#1ZZat\endcsname}\fi\next}
\expandafter\gdef\csname atZZZZ\endcsname#1{\errhelp
 \expandafter=\csname athelpZ\endcsname
 \csname errZ\endcsname{Invalid use of \string@}}}}
\def\atdef@#1{\expandafter\def\csname #1@at\endcsname}
\def\atdef@@#1{\expandafter\def\csname \string#1@@at\endcsname}
\newhelp\defahelp@{If you typed \string\define\space cs instead of
\string\define\string\cs\space^^J
I've substituted an inaccessible control sequence so that your^^J
definition will be completed without mixing me up too badly.^^J
If you typed \string\define{\string\cs} the inaccessible control sequence^^J
was defined to be \string\cs, and the rest of your^^J
definition appears as input.}
\newhelp\defbhelp@{I've ignored your definition, because it might^^J
conflict with other uses that are important to me.}
\atdef@.{\cd.}  
\atdef@>#1>#2>{\cd\rightarrow{#1}{#2}}
\atdef@<#1<#2<{\cd\leftarrow{#1}{#2}}
\atdef@ A#1A#2A{\cd\uparrow{#1}{#2}}
\atdef@ V#1V#2V{\cd\downarrow{#1}{#2}}
\atdef@|{\Big\Vert&&}
\atdef@@\vert{\Big\Vert&&}
\atdef@={&\hskip.5em\mathrel
 {\vbox{\hrule width\minCDaw@\vskip3\ex@\hrule width
 \minCDaw@}}\hskip.5em&}
\makeatother
\documentstyle[12pt]{article}

\font\sb=cmssdc10 scaled 1000
\begin{document}
\noindent {\large \bf Universal Formats for Nonlinear 
Dynamical Systems}\\

\vspace{1.5cm}
\noindent Krzysztof Kowalski\\

\vspace{1.5cm}
\noindent {\it Department of Theoretical Physics, University of \L\'od\'z,
ul.\ Pomorska 149/153,\\ 90-236 \L\'od\'z, Poland}
\vspace{1.5cm}
\begin{abstract}
It is demonstrated that very general nonlinear dynamical
systems covering all cases arising in practice can be
brought down to rate equations of chemical kinetics
\end{abstract}
\vspace{1.5cm}
\noindent Key words:\qquad \parbox[t]{12cm}{nonlinear
dynamical systems, chemical kinetics, chemical rate 
equations, graphs of chemical reactions}\\

\vspace{1.5cm}
\noindent PACS numbers:\qquad 02.10, 02.90, 82.20, 82.30
\newpage
In recent years there has been great progress in the theory 
of classical dynamical systems of the form\vspace{.4cm}
\begin{equation}
\dot x_i = F_i(x_1,\ldots ,x_k),\qquad i=1,\ldots ,k,
\end{equation}

\vspace{.4cm}
\noindent where overdot denotes differentiation with respect to the 
time.  Let us only recall the spectacular discovery of
dynamical systems showing chaotic behaviour [1].  On the
other hand, this part of the theory which deals with the 
universal properties of systems (1) can hardly be called 
satisfactory one.  For example the Sixteenth Hilbert Problem
(the problem of systematically counting and locating the
limit cycle of polynomial systems on the plane) posed in 
1900 remains unsolved even for systems with quadratic 
nonlinearity [2].

In order to study universal properties of dynamical systems 
(1) one should first find some universal object connected 
with (1).  An attempt in this direction was made by Kerner 
[3] who showed that very general dynamical systems (1) 
embracing everything that arises in practice can be brought 
down to polynomial systems\vspace{.4cm}
\begin{equation}
\dot x_i = P_i(x_1,\ldots ,x_N),\qquad i=1,\ldots ,N,
\end{equation}

\vspace{.4cm}
\noindent where $P_i$ are polynomials in $x_1,\ldots ,x_N$,
by introducing suitable new variables into the original 
system (1).  The aim of this letter is to demonstrate that 
the polynomial systems (2) resulting from the original
nonlinear dynamical systems (1) can be furthermore reduced
to systems of rate equations of chemical kinetics by 
introducing appropriate additional variables.  Thus we show 
in the present paper that very general dynamical systems (1) 
are represented by graphs of the corresponding chemical 
reactions.

We first recall the algorithm for construction of chemical 
rate equations from a given reaction mechanism.  Let the 
species of the chemical reaction mechanism be $A_1,\ldots 
,A_N$.  Elementary reactions are given by the stoichiometric 
equations\vspace{.4cm}
\begin{equation}
\sum\limits_{i=1}^{N}m_{ri}A_i \:\cd\rightarrow {k_r}{}
\:\sum\limits_{i=1}^{N}n_{ri}A_i,\qquad r=1,\ldots ,M,
\end{equation}

\vspace{.4cm}
\noindent where $m_{ri}$, $n_{ri}$ are the stoichiometric
coefficients, $m_{ri},\: n_{ri}\in \hbox{\sb Z}_+$ (the set
of nonnegative integers) and $k_r$ is the rate constant; the
formal sums from (3) are usually called complexes.
According to the mass action law, the concentrations $x_i$
of the species $A_i,\:i=1,\ldots ,N$\/ satisfy the
following polynomial system\vspace{.4cm}
\begin{equation}
\dot x_i =
\sum\limits_{r=1}^{M}(n_{ri}-m_{ri})k_r\prod\limits_{j=1}^{N
}x_j^{m_{rj}},\qquad i=1,\ldots ,N.
\end{equation}

\vspace{.4cm}
\noindent Consider now the general polynomial system (2).
One finds easily that it can be written in the
form\vspace{.4cm}
\begin{equation}
\dot x_i =
\sum\limits_{r=1}^{I_{i}}\alpha_{ir}\prod\limits_{j=1}^{N}
x_j^{n_{irj}} - 
\sum\limits_{s=1}^{J_i}\beta_{is}\prod\limits_{j=1}^{N}
x_j^{m_{isj}},\qquad i=1,\ldots ,N,
\end{equation}

\vspace{.4cm}
\noindent where $\alpha_{ir},\:\beta_{is}\in \hbox{\sb R}_+$ 
(the set of positive real numbers), the vectors ${\bf 
n}_{ir},\:{\bf m}_{is}\in \hbox{\sb Z}_+^N$ with coordinates 
$n_{irj}$ and $m_{isj}$, respectively are pairwise different 
and $I_p,\:J_q\in\hbox{\sb Z}_+$ (it is understood that if 
$I_p$ or $J_q = 0$ then the corresponding sum vanishes).  
The criterion for the polynomial system (5) to have an 
underlying chemical reaction mechanism has been found by 
H\'ars and T\'oth [4].  Namely, they observed that whenever 
the condition\vspace{.4cm}
\begin{equation}
m_{isi} > 0\quad \hbox{if}\quad J_i > 0,\qquad s=1,\ldots
,J_i
\end{equation}

\vspace{.4cm}
\noindent holds for arbitrary $i=1,\ldots ,N$, then (5) is
the system of rate equations corresponding to the following
sequence of chemical reactions
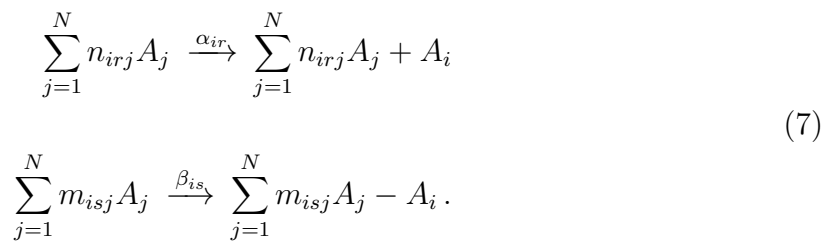
\vspace{.4cm}
\begin{eqnarray}
\sum\limits_{j=1}^{N}n_{irj}A_j \:\cd\rightarrow
{\alpha_{ir}}{} \:\sum\limits_{j=1}^{N}n_{irj}A_j +
A_i\nonumber \\
\null\\
\sum\limits_{j=1}^{N}m_{isj}A_j \:\cd\rightarrow
{\beta_{is}}{} \:\sum\limits_{j=1}^{N}m_{isj}A_j - A_i\,.\nonumber
\end{eqnarray}

\vspace{.4cm}
\noindent It can be easily checked that the general system
of rate equations (4) fulfils (6).  Therefore, (6) is the
necessary condition as well.  It should also be noted that
there can exist mechanisms different from (7) having the
same kinetic rate equations (5).  Nevertheless, the
advantage of (7) is that it can be constructed quickly and 
algorithmically.

We now demonstrate that the general polynomial system (5) 
can be easily reduced to this one satisfying the condition 
(6) by introducing suitable new coordinates.  Consider the 
system (5).  Suppose that the condition (6) is not valid for 
$i=i_1,\ldots ,i_q\in\{1,\ldots ,N\}$.  On introducing the
new variables of the form\vspace{.4cm}
\begin{equation}
x_{N+p} = 1/x_{i_p},\qquad p=1,\ldots ,q
\end{equation}

\vspace{.4cm}
\noindent we arrive at the following $N+q$--dimensional system
satisfying the condition (6)\vspace{.4cm}
$$\displaylines{\indent
\dot x_i =
\sum\limits_{r=1}^{I_i}\alpha_{ir}\prod\limits_{j=1}^{N}
x_j^{n_{irj}} -
\sum\limits_{s=1}^{J_i}\beta_{is}\prod\limits_{j=1}^{N}
x_j^{m_{isj}},\qquad i\in\{1,\ldots ,N\}\setminus \{i_1,\ldots
,i_q\},\hfill\cr
\indent \dot x_{i_p} =
\sum\limits_{k=1}^{I_{i_p}}\alpha_{i_pk}\prod\limits_{j=1}^{%
N}x_j^{n_{i_pkj}} -
x_{i_p}x_{N+p}\sum\limits_{l=1}^{J_{i_p}}\beta_{i_pl}\prod
\limits_{j=1}^{N}x_j^{m_{i_plj}},\hfill\llap{(9)}\cr
\indent\dot x_{N+p} = -x_{N+p}^2 \left(\sum\limits_{k=1}^{I_{i_p}}
\alpha_{i_pk}\prod\limits_{j=1}^{N}x_j^{n_{i_pkj}} -
\sum\limits_{l=1}^{J_{i_p}}\beta_{i_pl}\prod\limits_{j=1}^{N}
x_j^{m_{i_plj}}\right),\qquad p=1,\ldots ,q.\hfill\cr}$$

\vspace{.4cm}
\noindent We have thus shown that the general polynomial 
systems (5) can be brought down to these ones satisfying the 
condition (6) and the general dynamical systems (1) can be
reduced to rate equations of chemical kinetics.  An
alternative method for the reduction of polynomial systems
(2) to the chemical format has been recently reported by
Samardzija et al [5].  Nevertheless, this method seems to be
more complicated.  The system (9) can be furthermore reduced 
to quadratic one by successive grouping $x_jx_k,\:j,k=1,\ldots
,N+q$ as a single new variable [3].  Such reduction of
polynomial systems (5) to quadratic rate equations is the
ultimate one.  Indeed, if one persists in grouping $x_jx_k$
as new variables then the linear infinite dimensional system
is obtained (the nonlinear system (9) is embedded into
infinite linear system - such embedding is called the
Carleman one [6]).  The possibility of asymptotical
reduction of general polynomial systems (2) to quadratic
rate equations describing conservative chemical reactions
without autocatalytic steps was reported by Korzukhin [7].
Besides the perturbative nature the flaw of Korzukhin's
reduction is its complexity.  For example such simple
equation as $\dot x = x$ is reduced (asymptotically) within
the Korzukhin approach to rate equations corresponding to
chemical reactions of the above mentioned type by
introducing five new variables.\vspace{.8cm}

\noindent {\it Example}.\qquad Consider the Lorenz
system\vspace{.4cm}
$$\displaylines{%
\hbox{\hspace{5cm}}\dot x_1 = \sigma x_2 - \sigma x_1,\hfill\cr
\hbox{\hspace{5cm}}\dot x_2 = rx_1 - x_2 - x_1x_3,\hfill\llap{(10)}\cr
\hbox{\hspace{5cm}} \dot x_3 = x_1x_2 - bx_3\:.\hfill\cr}$$

\vspace{.4cm}
\noindent Owing to the term $\null-x_1x_3$ the system (10) is not
``chemical'' one.  Taking into account (8) we introduce a
new variable such that\vspace{.4cm}
$$x_4 = 1/x_2.$$

\vspace{.4cm}
\noindent The system (10) implies then the following system of rate
equations\vspace{.3cm}
$$\displaylines{%
\hbox{\hspace{4cm}}\dot x_1 = \sigma x_2 - \sigma
x_1,\hfill\cr
\hbox{\hspace{4cm}}\dot x_2 = rx_1 - x_2 -
x_1x_2x_3x_4,\hfill\cr}$$

\vspace{-.77cm}
\hfill(11)\vspace{-.77cm}
$$\displaylines{%
\hbox{\hspace{4cm}}\dot x_3 = x_1x_2 - bx_3,\hfill\cr
\hbox{\hspace{4cm}}\dot x_4 = x_4 + x_1x_3x_4^2 -
rx_1x_4^2\:.\hfill\cr}$$

\vspace{.4cm}
\noindent The sequence of chemical reactions (7)
corresponding to (11) is of the form\\[-2cm]
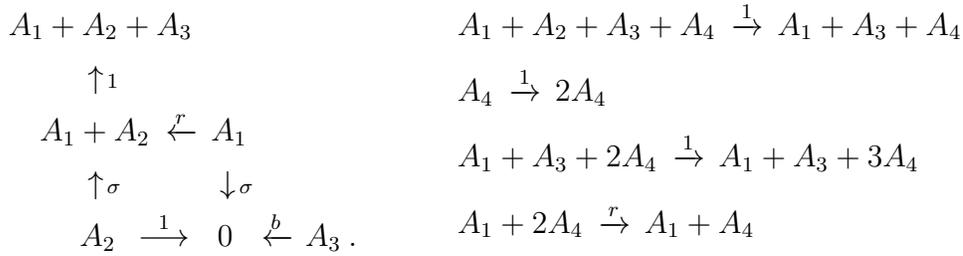
\begin{picture}(60,20)(-10,0)
\put(0,0){$A_1+A_2+A_3$}
\put(30,-20){$\uparrow\!\hbox{$\scriptstyle 1$}$}
\put(12,-40){$A_1+A_2 \:\cd\leftarrow {r} {}\: A_1$}
\put(30,-60){$\uparrow\!\hbox{$\scriptstyle\sigma$}$}
\put(80,-60){$\downarrow\!\hbox{$\scriptstyle\sigma$}$}
\put(27,-80){$A_2\:\:\stackrel{\hbox{$\scriptstyle 
1$}}{\longrightarrow}\:\:\,0\:\:\,\cd\leftarrow {b}{}\:A_3\:.$}
\put(170,0){$A_1+A_2+A_3+A_4\:\cd\rightarrow {1}{}\:A_1+A_3+A_4$}
\put(170,-25){$A_4\:\cd\rightarrow {1}{}\:2A_4$}
\put(170,-50){$A_1+A_3+2A_4\:\cd\rightarrow {1}{}\:A_1+A_3+3A_4$}
\put(170,-75){$A_1+2A_4\:\cd\rightarrow {r}{}\:A_1+A_4$}
\end{picture}

\vspace{3.5cm}
\noindent Here, 0 designates the zeroth complex and the reaction
$A\,\cd\rightarrow {}{}\,0$ ($0\,\cd\rightarrow {}{}\,A$) means that
the substance A is removed from (supplied to) the chemical
reactor.  In order to reduce (11) to the system of quadratic
rate equations we now introduce the following new
variables\vspace{.4cm}
\setcounter{equation}{11}
\begin{equation}
x_5 = x_1x_4,\qquad x_6 = x_3x_5.
\end{equation}

\vspace{.4cm}
\noindent Making use of (11) one finds easily that in this variables
the vector fields $\dot x_i$, where $i\neq 4$ do not depend
on $x_4$ and therefore the variable $x_4$ can be omitted from
the six-dimensional system obtained via (11) and (12).  On
renumbering $x_6\to x_4$ we finally arrive at the system of
quadratic rate equations such that\vspace{.4cm}
$$\displaylines{\hbox{\hspace{3cm}}\dot x_1 = \sigma x_2 -
\sigma x_1, \hfill\cr
\hbox{\hspace{3cm}}\dot x_2 = rx_1 - x_2 - x_2x_4, \hfill\cr
\hbox{\hspace{3cm}}\dot x_3 = x_1x_2 - bx_3, \hfill\cr
\hbox{\hspace{3cm}}\dot x_4 = \sigma x_3 + x_1^2 + x_4^2
-(\sigma + b - 1)x_4 - rx_4x_5, \hfill\cr
\hbox{\hspace{3cm}}\dot x_5 = \sigma + x_4x_5 - (\sigma -
1)x_5 -rx_5^2, \hfill\cr}$$

\vspace{.4cm}
\noindent where $x_4=x_1x_3/x_2,\:x_5=x_1/x_2$.  The
corresponding chemical reaction mechanism (7) can be
written as
\newpage
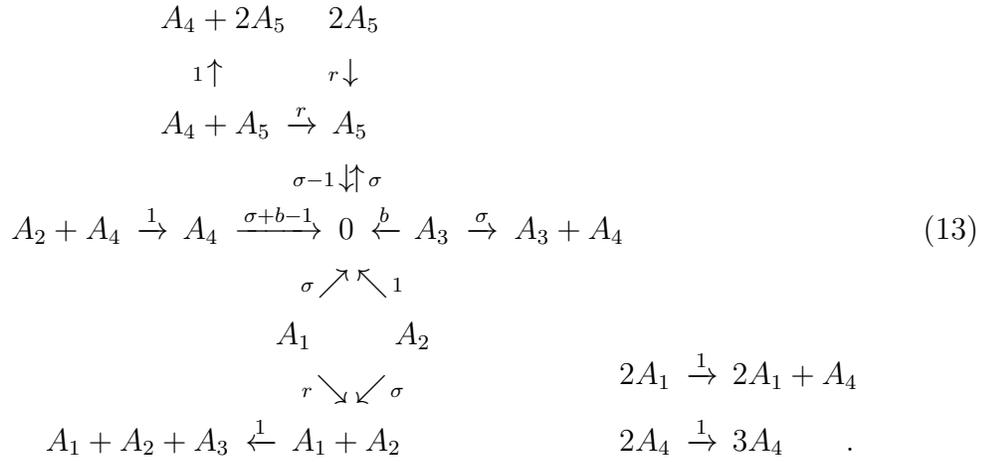
\begin{picture}(60,20)(-40,0)
\put(0,-80){$A_2+A_4\:\cd\rightarrow
{1}{}\:A_4\:\cd\rightarrow {\sigma + b
-1}{}\:0\:\cd\leftarrow {b}{}\:A_3\:\cd\rightarrow
{\sigma}{}\:A_3+A_4$}
\put(345,-80){(13)}
\put(106.5,-60){$\hbox{$\scriptstyle \sigma -1$}\!\downarrow\!\uparrow\!
\hbox{$\scriptstyle \sigma$}$}
\put(56.5,-40){$A_4+A_5\:\cd\rightarrow {r}{}\:A_5$}
\put(68.5,-20){$\hbox{$\scriptstyle 1$}\!\uparrow$}
\put(120,-20){$\hbox{$\scriptstyle r$}\!\downarrow$}
\put(56.5,0){$A_4+2A_5$}
\put(120,0){$2A_5$}
\put(109.5,-100){$\hbox{$\scriptstyle \sigma$}\!\nearrow\:
\nwarrow\!\hbox{$\scriptstyle 1$}$}
\put(100,-120){$A_1$}
\put(145,-120){$A_2$}
\put(110,-140){$\hbox{$\scriptstyle r$}\!\searrow\:
\swarrow\!\hbox{$\scriptstyle \sigma$}$}
\put(13,-160){$A_1+A_2+A_3\:\cd\leftarrow {1}{}\:A_1+A_2$}
\put(230,-160){$2A_4\:\cd\rightarrow {1}{}\:3A_4$\qquad .}
\put(230,-135){$2A_1\:\cd\rightarrow {1}{}\:2A_1+A_4$}
\end{picture}

\vspace{8cm}
In summary, we have demonstrated in this work that very
general classical dynamical systems can be reduced to rate
equations of chemical kinetics.  The cost of such reduction
is an increase in dimension of systems resulting from
introducing new variables and restrictions imposed on
initial data by the requirement of nonnegativity of
concentrations.  In spite of these limitations, the point
remains that general dynamical systems can be viewed as
kinetic rate systems and thus they are represented by graphs
of the corresponding chemical reactions.  The problem of
possible applications of this result remains open.  On the
one hand, working in the field of nonlinear models will find
that existing surprisingly strong theorems relating the
dynamics of rate equations with the structure of the
underlying graph of a chemical reaction mechanism such as
zero defficiency theorem [8], the theorem on knots of trees
[9] and Volpert theorems [10] (see also [11]) cannot be used
in the case of graphs such as (13) corresponding to reaction
mechanisms with autocatalytic steps.  On the other hand, as
it was mentioned by a referee, a chemist is most interested
in inverse problem: ``how to find a simple model underlying
the complex reality of a typical chemical mechanism'' (in
the light of the observations presented herein it is really
difficult job).  Nevertheless, maybe at some distant stage
in the future our experimental ability will be such that it
would be possible to ``design'' particular chemical reactor
that fit given scheme and, for example, one would predict
behaviour of a complex mechanical system from the colour of
a chemical mixture.

\vspace{.8cm}
\noindent {\bf Acknowledgement}

\vspace{.4cm}
I would like to thank a referee for helpful comments.
\newpage
\noindent {\bf References}

\vspace{.8cm}
\noindent [1]\quad \parbox[t]{14cm}{E.N. Lorenz, J.~Atmos.~
Sci. {\bf 20} (1972) 130.}\\

\noindent [2] \quad \parbox[t]{14cm}{D.E. Koditschek and
K.S. Narendra, J.~Differential Equations {\bf 54} (1984)
181.}\\

\noindent [3] \quad \parbox[t]{14cm}{E.H. Kerner,
J.~Math.~Phys. {\bf 22} (1981) 1366.
}\\

\noindent [4] \quad \parbox[t]{14cm}{V. H\'ars and J.
T\'oth, in:\quad Qualitative Theory of Differential
Equations, Szeged, 1979, Colloq.~Math.~Soc. J\'anos Bolyai
{\bf 30}.}\\[.5cm]

\noindent [5] \quad \parbox[t]{14cm}{N. Samardzija, L.D.
Geller and E. Wasserman, J.~Chem.~Phys. {\bf 90} (1989)
2296.}\\[.5cm]

\noindent [6] \quad \parbox[t]{14cm}{K. Kowalski and W.-H.
Steeb, Nonlinear Dynamical Systems and Carleman
Linearization (World Scientific, Singapore, 1991).}\\[.5cm]

\noindent [7] \quad \parbox[t]{14cm}{M.D. Korzukhin,
in:\quad Oscillatory Processes in Biological and Chemical
Systems (Nauka, Moscow, 1967) (in Russian).}\\[.5cm]

\noindent [8] \quad \parbox[t]{14cm}{M. Feinberg, in:\quad
Dynamics and Modelling of Reactive Systems (Academic Press,
New York, 1980).}\\[.5cm]

\noindent [9] \quad \parbox[t]{14cm}{B.L. Clarke,
Adv.~Chem.~Phys. {\bf 43} (1980) 7.}\\

\noindent [10] \quad \parbox[t]{14cm}{A.I. Volpert and S.I.
Khudayev, Analysis in Classes of Discontinuous Functions and
Equations of Mathematical Physics (Nauka, Moscow, 1975) (in
Russian).}\\[.5cm]

\noindent [11] \quad \parbox[t]{14cm}{A.N. Gorban, V.I.
Bykov and G.S. Yablonski, Essays on Chemical Relaxation
(Nauka, Novosibirsk, 1986) (in Russian).}
\end{document}